\documentclass[acus]{JAC2001}

\usepackage{graphicx}

\begin{document}

\title{\flushright{T24}\\[15pt]
\centerline{ Production at Intermediate Energies and Lund Area
Law}}

\author{Haiming Hu\thanks{Institute of High Energy Physics,
Academia Sinica, Beijing 10039, China}, An Tai\thanks{Department
of Physics and Astronomy, University of California, Los Angeles,
CA90095, USA}}

\maketitle

\begin{abstract}

The Lund area law was developed into a Monte Carlo program LUARLW.
The important ingredients of this generator was described. It was
found that the LUARLW simulations are in good agreement with the
BEPC/BES $R$ scan data between 2--5 GeV.
\end{abstract}

\section{Introduction}

\begin{figure*}[tb]
\includegraphics[width=14cm,height=20cm]{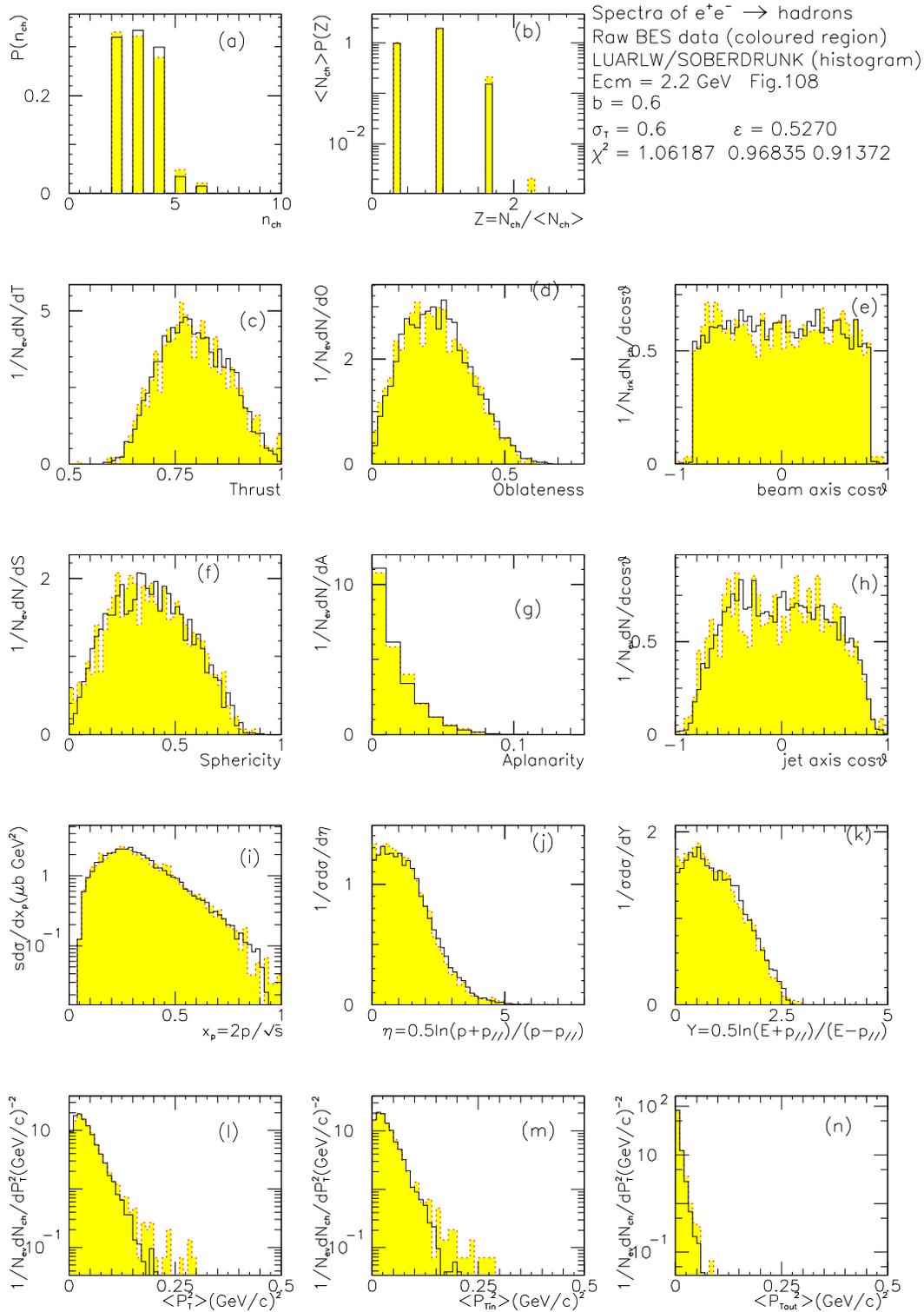}
\caption{\label{fig2200}$e^+e^-\rightarrow$ hadrons spectrum of
         raw BES data (hatched region) and LUARLW/SOBDRUNK
         (black line) at $Ecm = 2.2$ GeV.}
\end{figure*}

The hadron production mechanism in particles collisions is one of
the important subjects in the study of strong interaction. Quantum
chromodynamics (QCD) is considered as the theory of strong
interaction. However, the hadronization processes belong to
nonperturbative problem for which no practicable calculation based
on the first principle available. Some phenomenological
hadronization models were thus built up, which play important
roles in the studies toward the final understanding of strong
interaction. The famous Lund string fragmentation model is one of
the successful hadronization schemes, which contains several
nontrivial dynamical features and describes the general
semi-classical picture of hadron production. At high energies, the
Lund generator, JETSET, can simulate the processes of hadron
production via single photon annihilation and predicts the many
properties of the final states correctly. But the application of
Lund model at intermediate energies has been blank. A direct way
out of this situation is to start from the basic assumptions of
Lund model and find the solutions of the area law without adopting
any high-energy approximation. Based on the Lund area law, a new
generator LUARLW was compiled, which agrees with BES data between
$2-5$ GeV well (see Figure \ref{fig2200} on page
\pageref{fig2200}).

\section{Lund string fragmentation}

\begin{figure}[tb]
\centering
\includegraphics[width=\columnwidth]{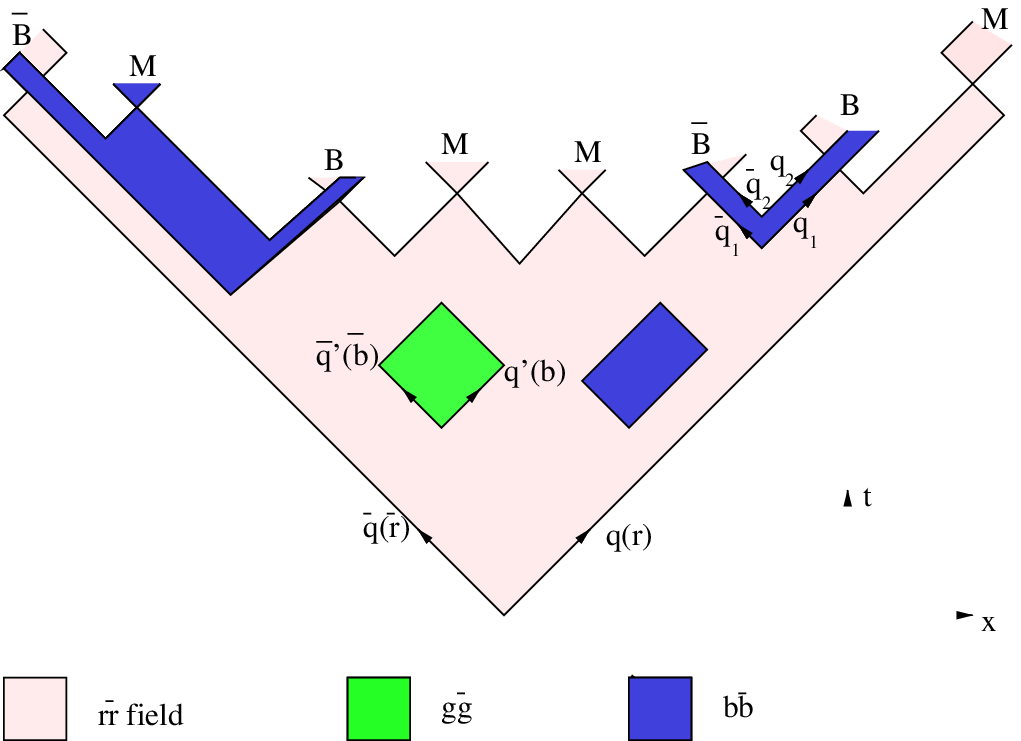}

\caption{String fragmentation by a set of new pairs $(q\bar{q})$
and $(q\bar{q}q\bar{q})$ production, hadrons form at
vertices\label{haiming:breakup}}
\end{figure}

The foundations of Lund model (relativity, causality and quantum
mechanics) are universal. The basic hadron production picture is
string fragmentation. The produced new pairs $(q\bar{q})$ and
$(q\bar{q}q\bar{q})$ may form mesons and baryons if they carry
with the correct flavor quantum numbers, otherwise they just
behave like the vacuum fluctuations and do not lead any observable
effects in experiments (see Figure \ref{haiming:breakup}).
Using the assumptions of very high energy approximation (the
remaining string always has large energy scale), left--right
symmetry (fragmentation from $q_0$ end or $\bar{q}_0$ end are
identical) and iterative fragmentation (string fragmentation may
be treated iteratively), Lund fragmentation function $f(z)$ was
derived uniquely,
\begin{equation}
     f(z)=\frac{N}{z}(1-z)^a \exp\left(-\frac{bm^2}{z}\right),
\end{equation}
where, $a$ and $b$ are fragmentation parameters, $m$ is the
(transverse) mass of fragmentation hadron, and $z$ is the fraction
of light-cone momentum. $f(z)$ is used in JETSET to govern string
fragmentation. Lund fragmentation function $f(z)$ has the
characteristics of inclusive distribution, and the single particle
production is independent of anything else before and after. The
applicable region of $f(z)$ is the remnant string still has large
invariant mass. At intermediate energies, the mass-shell
conditions should be the component part of the fragmentation
dynamics, and the string usually fragments into $2\sim 6$ hadrons.
Therefore the string fragmentation have to be treated as exclusive
one instead of inclusive like in JETSET.

\section{Lund area law}

\begin{figure}[tb]
\includegraphics[width=\columnwidth]{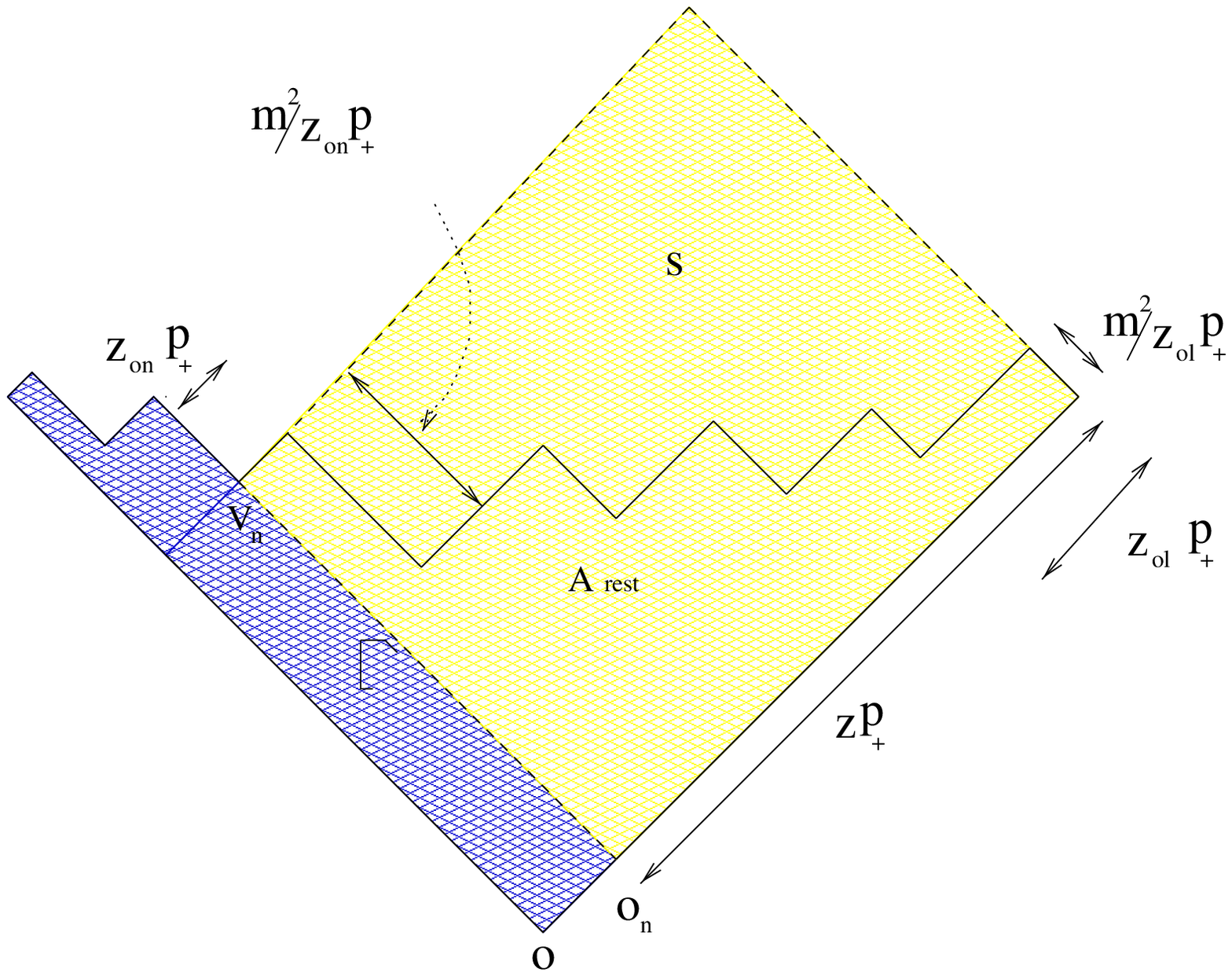}

\caption{\label{cluster}The situation after $n$ steps
fragmentation.}
\end{figure}

Lund string fragmentation process is Lorentz invariant and
factorizable. The finite energy ($s$) system containing $n$
hadrons may be viewed as a cluster of infinite string
fragmentation system with energy ($s_0\to \infty$) (see Figure
\ref{cluster}). According to the general properties of iterative
cascade, the combined distribution is the product of fragmentation
functions for every steps
\begin{eqnarray}\label{e2b4}
     d\tilde{\wp}_n &=&\prod _{j=1}^{n}f_j(z_j)dz_j\nonumber\\
     &= &C_n\cdot d\wp_\mathrm{ext}(s,z)\cdot d\wp_{n}(u_1,\cdots ,u_n),
\end{eqnarray}
where, $C_n$ is normalization constant. We know from (\ref{e2b4})
that a subsystem may be split up from the total system, the
processes occurring in the subsystem is the same as it be a
complete system starting at the some original energy $s$. The
external part
\begin{equation}\label{e2b10}
     d\wp_\mathrm{ext}(s,z)=ds\frac{dz}{z}(1-z)^a\cdot \exp(-b\Gamma )
\end{equation}
corresponds to the probability that the cluster will occur. The
internal part
\begin{eqnarray}\label{arealaw}
     &&d\wp_{n}(u_1,\ldots ,u_n)
     =\delta^2\left(P_n-\sum _{j=1}^{n}p_{\circ j}\right)\nonumber\\
    &&\qquad \cdot\prod _{j=1}^{n}d^2p_{\circ j}\delta
     (p_{\circ j}^2-m_{\perp j}^2)\exp(-b\mathcal{A}_{n})
\end{eqnarray}
then corresponds to the exclusive probability that the cluster
will decay into the particular channel containing the given $n$
particles with energy-momentum $\{p_{\circ j}\}$ and nothing else.
$\mathcal{A}_\mathrm{rest}=\mathcal{A}_n$ is the area enclosed by
the quark and antiquark light-cone energy-momentum lines of $n$
particles.
The factor
\begin{equation}
     |M|^2\equiv \exp(-b \mathcal{A}_{n})
\end{equation}
may be viewed as the squared matrix element, the other parts are
phase-space elements. In formulas, $b$ is fundamental
color-dynamical parameters. Distribution (\ref{arealaw}) is called
Lund area law. The total area
\begin{equation}\label{e2b6}
     \mathcal{A}_\mathrm{tot}=\sum _{j=1}^{n}\mathcal{A}_j
     = \mathcal{A}_{n}+\Gamma,\qquad \Gamma=\frac{s(1-z)}{z}
\end{equation}
and
\begin{equation}\label{e2b8}
     \mathcal{A}_{n}=\sum _{j=1}^{n}\frac{m_{\perp j}^2}{z_j}\cdot
     \left(\sum _{k=j}^{n}z_k\right).
\end{equation}
Finishing the integral over kinematic variables of $n-$particles,
area law has following forms:
\begin{itemize}
\item String $\Rightarrow$ 2 hadrons
\begin{equation}
     \wp_2=\frac{C_2}{\sqrt{\lambda}}\left[\exp(-b\mathcal{A}_2^{(1)})
     +\exp(-b\mathcal{A}_2^{(2)})\right].
\end{equation}

\item String $\Rightarrow$ 3 hadrons
\begin{equation}
     d\wp_3=\frac{C_3}{\sqrt{\Lambda}}\exp(-b\mathcal{A})d
     \mathcal{A},
\end{equation}

\item String $\Rightarrow 4,5,6$ hadrons
\begin{eqnarray}
     d\wp_n(s)&= &\frac{ds_1ds_2}{\sqrt{\lambda (s,s_1,s_2)}}
     \exp(-b\Gamma )\nonumber\\
     &\cdot&\wp_{n_1}(s_1,\mathcal{A}_1)\wp_{n_2}
     (s_2,\mathcal{A}_2).
\end{eqnarray}
\end{itemize}

In above fragmentation distributions, the gluon effects are
neglected. At intermediate and low energies, the emitted gluons
from initial quark or antiquark are usually soft, most of which
will stop before the string starts to break, the effect of the
gluon will then essentially be small transverse broadening of
two-jet system, the gluon and quark will then look as single quark
jet. The gluon emissions do not significantly change the
topological shapes (sphericity and thrust) of final states, and
therefore no observable jet effects.

\begin{figure}[tb]
\centering
\includegraphics[height=\columnwidth,angle=-90]{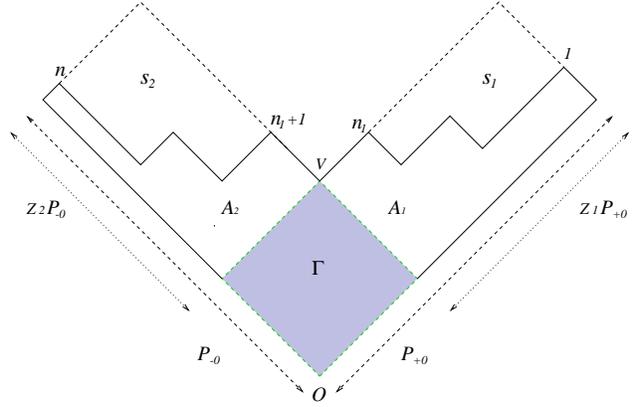}

\caption{\label{mlbodyarea} The vertex $V$ divides the $n$-body
string fragmentation into two clusters which contain $n_1$ and
$n_2$ hadrons and with squared invariant masses $s_1$ and $s_2$
separately.}
\end{figure}

\section{Multiplicity}

Lund area law may give the expression of the multiplicity
distribution of fragmental hadrons. Define dimensionless
$n$-particle partition function
\begin{equation}
     Z_n=s\int dR_n\cdot \exp(-b\mathcal{A}),
\end{equation}
where $dR_n$ is the $n$-particle phase space element. The relation
between $Z_n$ and the multiplicity distribution $\tilde{P}_n$ for
primary hadrons is
\begin{equation}
     \tilde{P}_n=\frac{Z_n}{\sum Z_n}.
\end{equation}
$\tilde{P}_n$ has the approximative expression
\begin{eqnarray}
     \tilde{P}_n &= &\frac{\mu^n}{n!}\cdot \exp[c_0+c_1(n-\mu)\nonumber\\
     &&\qquad +c_2(n-\mu)^2],
     \qquad (c_2 <0).
\end{eqnarray}
Quantity $\mu$ may written as the energy-dependent form
phenomenologicaly
\begin{equation}
     \mu=a+b\cdot \exp(c\sqrt{s}),
\end{equation}
or
\begin{equation}
     \mu=a+b \ln(s)+c \ln^2(s).
\end{equation}
All parameters $a$, $b$, $c$, $c_0$, $c_1$, and $c_2$ need to be
determined by experimental data and have been tunned with BES data
samples of $R$ scan.

\section{Exclusive distribution}

There are some different production channels for $n$-particle
states, such as 4-body states may be $\pi^+\pi^-\pi^+\pi^-$,
$\pi^+\pi^-\pi^{\circ}\pi^{\circ}$, $\rho^+\rho^-\pi+\pi^-$, etc.
The exclusive probability for the special channel is
\begin{equation}\label{probn}
     \hat{P}_n=B_n\cdot (VPS)\cdot (SUD)\cdot
     \wp_n(m_{\perp 1},\ldots,m_{\perp n};s).
\end{equation}
\begin{itemize}
\item $B_n$ is the combinatorial number stemming
from may be more than one string configurations lead to this
state.

\item (VPS) is the vector to pseudoscalar rate.

\item (SUD) is the strange to up and down quark
pair probability.
\end{itemize}

\section{Transverse momentum distribution}

Above results are obtained when the transverse momentums of all
primary hadrons have given. In LUARLW, two transverse momentum
distributions were used alternatively.

\subsection{Scheme I}

In Lund model, quantum mechanical tunneling effect is was used to
explain the production of new pairs $q_i\bar{q}_i$. Particles
obtain their transverse momenta from the constituents. At each
production point the $(q_i\bar{q}_i)$-pair is given $\pm{\bf q}_i$
and the particle momenta are
\begin{equation}
     \mathbf{p}_{\perp 1}=\mathbf{q}_1,\ldots
     \mathbf{p}_{\perp j}=\mathbf{q}_j-\mathbf{q}_{j-1},
     \ldots \mathbf{p}_{\perp n}=-\mathbf{q}_n.
\end{equation}
Based on the Lund model, the following distribution with
forward--backward symmetric correlation was derived
\begin{eqnarray*}
     &&\hspace{-20pt}F^{(n)}(\mathbf{q}_1,\ldots ,\mathbf{q}_n)\\
     &&=C_n \exp\left\{ -\frac{1}{2\sigma^2}
     \left[\mathbf{q}_1^2+\frac{(\mathbf{q}_2-\rho_2
     \mathbf{q}_1)^2}{1-\rho_2^2}\cdots \right]\right\}\\
     &&=C_n \exp\left\{-\frac{1}{4\sigma^2}\left[\mathbf{q}_1^2
     + \mathbf{q}_n^2\vphantom{\sum}\right.\right.\\
     &&\qquad \left.\left. +\sum A_j(\mathbf{q}_j^2 +\mathbf{q}_{j-1}^2
     -2 \varepsilon_j\mathbf{q}_j\cdot\mathbf{q}_{j-1})\right]
     \vphantom{\frac{1}{4\sigma^2}}\right\},
\end{eqnarray*}
with
\begin{equation}
     A_j= \frac{(1+\rho_j^2)}{(1-\rho_j^2)},\qquad
     \varepsilon_j= \frac{2 \rho_j}{(1+\rho_j^2)}.
\end{equation}
The correlations $\rho_j$ are phenomenological parameters, which
in general are small. The covariant matrices
\begin{eqnarray*}
     V_2 &= &\sigma^2\left(\begin{array}{cc}
                       1      & \rho_2 \\
                       \rho_2 &   1
                       \end{array} \right),\\
     V_3 &= &\sigma^2\left(\begin{array}{ccc}
                        1       & \rho_2 & \rho_2\rho_3\\
                        \rho_2  &   1    & \rho_3 \\
                        \rho_2\rho_3 & \rho_3 & 1
                           \end{array} \right),\\
     V_n &= &\\
     &&\hspace{-20pt}\sigma^2\left(\begin{array}{cccc}
                1 & \rho_2 &  \cdots & \rho_2\rho_3\cdots \rho_n \\
                \rho_2 & 1 &  \cdots & \rho_3\cdots \rho_n      \\
                \cdots  & \cdots  & \cdots & \cdots \\
                \rho_2\cdots\rho_{n-1} & \rho_3\cdots\rho_{n-1}  &
                \cdots & \rho_n \\
                \rho_2\cdots\rho_n & \rho_3\cdots\rho_n &\cdots &  1
                 \end{array} \right)
\end{eqnarray*}
give the correlations $\langle p_{x i}p_{x j}\rangle$ and $\langle
p_{y i}p_{y j}\rangle$.

\subsection{Scheme II}

An available Gaussian-like transverse momentums distribution as
options in LUARLW for $n-$particles fragmentation reads
\begin{eqnarray}
     F^{(n)}(\mathbf{p}_{\perp 1},\ldots ,\mathbf{p}_{\perp n})
      &=&\delta \left(\sum\limits_{j=1}^{n}
      \mathbf{p}_{\perp j}\right)\nonumber\\ [3pt]
     &&\hspace{-4pc}\theta \left(\sqrt{s}-\sum _{j=1}^{n}
     \sqrt{m_j^2+ \mathbf{p}_{\perp j}^2}\right)\nonumber\\ [3pt]
     &&\hspace{-4pc}\cdot\prod _{j=1}^{n}\exp\left(-\frac{\mathbf{p}_{\perp
     j}^2}{2\sigma^2}\right).
\end{eqnarray}
The conditional distribution of $\mathbf{p}_{\perp 1}$,
$\mathbf{p}_{\perp 2}, \ldots, \mathbf{p}_{\perp j}$ are
\begin{eqnarray*}
     f_1^{(n)}(\mathbf{p}_{\perp 1})
         &=&N_1\cdot \exp\left[-\frac{n}{n-1}
          \frac{\mathbf{p}_{\perp 1}^2}{2\sigma^2}\right],\\[5pt]
     f_2^{(n)}(\mathbf{p}_{\perp 2})
         &=&N_2 \exp\left[-\frac{n-1}{n-2}
          \frac{(\mathbf{p}_{\perp 2}
          +\frac{\mathbf{p}_{\perp
          1}}{n-1})^2}{2\sigma^2}\right],\\[5pt]
     f_j^{(n)}(\mathbf{p}_{\perp j})&=&\\
     &&\hspace{-3pc}N_j \exp\left[-\frac{(n-j+1)}{(n-j)}
          \frac{(\mathbf{p}_{\perp j}+\sum _{i=1}^{j-1}
          \frac{\mathbf{p}_{\perp j}}{n-j+1})^2}{2\sigma^2}\right].
\end{eqnarray*}
The final $\mathbf{p}_{\perp n}$ is determined by energy-momentum
conservation.The effective variance of $\mathbf{p}_{\perp j}$ is
\begin{equation}
     \sigma_{j}^\mathrm{(eff)}=\sqrt{\frac{n-j}{n-j+1}} \sigma,
     \qquad (j=1,\ldots,n-1).
\end{equation}
The transverse momentum distribution in this scheme is not exact
Gaussian type due to the transverse momentum conservation and the
threshold conditions.

\section{Summary}
The well-know Monte Carlo simulation packet JETSET is not built in
order to describe few-body states at the few GeV level in $e^+e^-$
annihilation as at BEPC. We develop the formalism to use the basic
Lund Model area law directly for Monte Carlo program LUARLW, which
will be satisfied to treat two-body up to six-body states. In
LUARLW, the effects of all gluonic emissions were neglected. The
LUARLW predicts more than 14 distributions totally agree with BES
data well.

\paragraph{Acknowledgement} This work was done under the instruction by
Prof. Bo Andersson. All the knowledge about Lund model may be fond
in The Lund Model by Bo Andersson, Cambridge University Press,
Cambridge, 1998.

\end{document}